\documentclass[twocolumn,amsmath,amssymb,aps]{revtex4}%
\usepackage{graphicx}
\usepackage{bm}
\usepackage{amsmath}
\usepackage{amsfonts}
\usepackage{amssymb}%
\newcommand{\be}{
\begin{equation}
}
\newcommand{\ee}{
\end{equation}
}
\newcommand{\beq}{
\begin{eqnarray}
}
\newcommand{\eeq}{
\end{eqnarray}
}

\begin{document}
\title{Searching fast for a target on a DNA without falling to traps}
\author{O. B\'enichou$^{1}$, Y. Kafri$^{2}$, M. Sheinman$^{2}$ and R. Voituriez$^{1}$}
\affiliation{$1$ UMR 7600, Universit\'e Pierre et Marie Curie/CNRS, 4 Place Jussieu, 75255
Paris Cedex 05 France.}
\affiliation{$2$ Department of Physics, Technion, Haifa 32000, Israel.}
\date{\today}

\begin{abstract}
Genomic expression depends critically both on the ability of
regulatory proteins to locate specific target sites on a DNA within
seconds and on the formation of long lived (many minutes) complexes
between these proteins and the DNA. Equilibrium experiments show
that indeed regulatory proteins bind tightly to their target site.
However, they also find strong binding to other non-specific sites
which act as traps that can dramatically increase the time needed to
locate the target. This gives rise to a conflict between the speed
and stability requirements. Here we suggest a simple mechanism which
can resolve this long-standing paradox by allowing the target sites
to be located by proteins within short time scales even in the
presence of traps. Our theoretical analysis shows that the mechanism
is robust in the presence of generic disorder in the DNA sequence
and does not require a specially designed target site.
\end{abstract}
\date{\today}

\maketitle

It is commonly believed that three-dimensional diffusion is too slow for
proteins to locate their specific target on a DNA molecule for cells to
function properly. To resolve this issue Berg and von Hippel suggested, in
series of seminal papers \cite{WBH81,HB89}, that combining periods of
one-dimensional diffusion along the DNA (sliding) with periods of
three-dimensional diffusion off the DNA (jumping)
can speed up the search time by several orders of magnitude. Since then,
sliding (or equivalently binding of proteins to non-specific DNA sequences)
has been observed in many experiments \cite{WAC2006,Elf2007wq,Bonnet2008rp}
and is now believed to be a common mechanism
\cite{SM2004,nousprotein,Grosberg,Lomholt2005,Eliazar2007,Broek2008kl}. On the
other hand, as pointed out already in \cite{BH87}, experimental and
theoretical works have shown that the binding energies of a protein to
different DNA sequences are very large - a direct consequence of the required
stability of the protein with its target site. The binding energies can be
well fitted by a Gaussian with the strongest binding energies of the order of
$\sim30 k_{B}T$ and a standard deviation of the order of $5 k_{B}T$
\cite{GMH2002}. This casts a cloud on the simple facilitated diffusion picture
of Berg and von Hippel - the binding energy distribution suggests an
unacceptably slow search with very slow sliding and deep traps
\cite{SM2004}. This unresolved conflict is called the
\textit{speed-stability paradox} \cite{WBH81}.

\begin{figure}[ptb]
\begin{center}
\includegraphics[width=6.5cm]{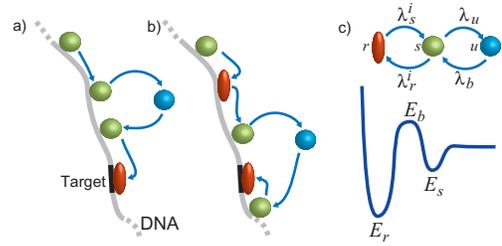}
\end{center}
\caption{An illustration of the model. (a) A time sequence of a
protein sliding in the $s$ mode (green circle), diffusing off the
DNA (blue circle) and entering the target site in the $r$ mode (red
oval). (b) A protein finding the target after entering the $r$
state. (c) An illustration of the rates and the energy landscape
which governs them at each location, $i=1,...,N$, along the DNA.
Here $\lambda^{i}_{r} \propto e^{-(E^{i}_{b}-E^{i}_{s})/k_{B}T}$,
$\lambda ^{i}_{s} \propto e^{-(E^{i}_{b}-E^{i}_{r})/k_{B} T}$ and
$\lambda_{u} \propto e^{-E^{i}_{s}/k_{B}T}$, while $\lambda_{b}$
depends on details of the three-dimensional diffusion process.}
\label{fig1}%
\end{figure}

Here, motivated by direct experimental observations
\cite{KBBLGBK2004,PW2007,TSBXA2007} and the theoretical work by Slutsky and
Mirny \cite{SM2004}, we consider a model in which the protein, when bound to
the DNA, can switch between two conformations separated by a free energy barrier.
In one, termed the search state the protein is loosely bound to the DNA and
can slide along it. In the second, recognition mode, it is trapped in a deep
energy well. Note that equilibrium measurements of binding energies to the DNA
are controlled by the recognition state.

In this paper, based on a quantitative analysis of this model, we argue that
due to the occurrence of several time scales in the search process the widely
used definition of the reaction rate of a single protein as the inverse of the
average search time $t^{ave}$ \cite{HANGGI1990}, is generally irrelevant as a
measure of the efficiency of target location on DNA. When
$n_{p}$ proteins are searching for the target, the relevant quantity is the
probability $\mathcal{R}_{n_{p}}(t)$ for a reaction to occur before time $t$.
We  show below that $\mathcal{R}_{n_{p}}(t)$ can reach values
close to one in a time scale $t^{typ}_{n_{p}}(t)$ which can be orders of
magnitude smaller than the value $t^{ave}/n_{p}$ expected from the usual approach.

\begin{figure}[ptb]
\begin{center}
\includegraphics[width=6.5cm]{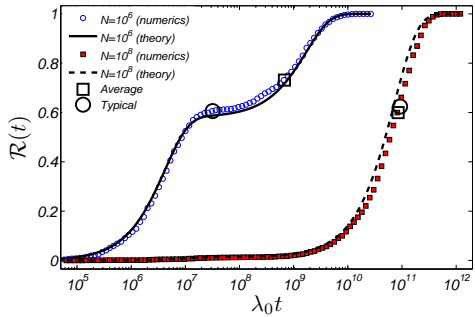}
\end{center}
\caption{A plot of $\mathcal{R}(t)$ for $N=10^{6}$ (empty circles)
and $N=10^{8}$ (filled squares). Lines correspond to Eq. \ref{2exp},
with $\tau_{1}$, $\tau_{2}$ and $q$ derived analytically. Here
$\lambda_{u}=10^{-2}\lambda_0$, $\lambda_{b}=0.1\lambda_0$
,$\lambda_{r}=10^{-7}\lambda_0$ and $\lambda_{s}=10^{-9}\lambda_0$.
These correspond to energies, measured relative to the $s$ mode, of
$E_{3}=4.6k_{B} T$, $E_{barrier}=16.12k_{B}T$ and
$E_{r}=-4.6k_{B}T$. Experiments suggest
$\lambda_{0} \simeq10^{6} \sec^{-1}$  for the Lac repressor \cite{WAC2006}.}
\label{fig2}%
\end{figure}

Our analysis has several important merits. First, it reports a \textit{fast}
search time despite a very strong binding of the protein in the recognition
state to \textit{any site} on the DNA. We suggest that the measured
binding energies of proteins to the DNA are irrelevant to the kinetics of the search process;
the relevant quantities are transition rates (specified below).
Second, it shows that in the realistic case of generic disorder in the barrier
height the search can be very effective even if the target site is \emph{not} designed. If experimentally verified the proposed mechanism will
resolve the speed-stability paradox.

The model consists of $n_{p}$ proteins which can each be in three
states (i) an unbound state, $u$, in which it performs
three-dimensional diffusion (jumping), (ii) a search state, $s$,
where it is weakly bound to the DNA, performing one-dimensional
diffusion (sliding) and (iii) a recognition state, $r$, where it is
tightly bound to the DNA. We assume, for simplicity, that in the
recognition state the protein is trapped in a deep energy well (as
justified by the experimentally measured strong binding energies)
and is unable to move \cite{SM2004}. The transition rates,
$\lambda_{s}^{i}$, $\lambda_{r}^{i}$, $\lambda_{b}$ and
$\lambda_{u}$, between the different states are defined in Fig.
\ref{fig1}. To model sliding, in the $s$-state the protein can move
with rate $\lambda_{0}/2$ to neighboring sites on the DNA. Note that
the rates $\lambda^{i}_{r}$ and $\lambda^{i}_{s}$ may depend on the
location $i=1 \ldots N$ along the DNA. In principle $\lambda_{0}$
and $\lambda_{u}$ also have a dependence on $i$. As justified later
this will have a weaker effect on our results and we omit it for
clarity. Finally, after a jump we assume the protein relocates to a
random position on the DNA due to its packed conformation
\cite{SK2009}.

\begin{figure}[ptb]
\begin{center}
\includegraphics[width=6.5cm ]{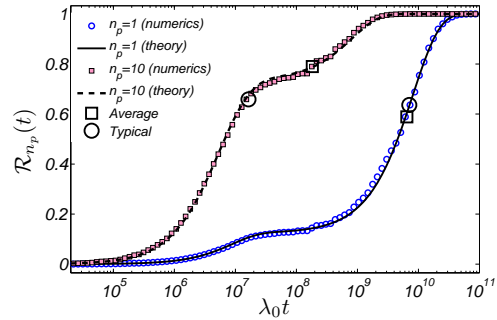}
\end{center}
\caption{A plot of $\mathcal{R}_{n_{p}}(t)$ for $n_{p}=1$ (empty
circles) and $n_{p}=10$ (filled squares). Here $N=10^{6}$,
$\lambda_{u}=10^{-4}\lambda_0$, $\lambda_{b}=0.1\lambda_0$,
$\lambda_{r}=10^{-7}\lambda_0$ and $\lambda_{s}=10^{-9}\lambda_0$.
These correspond to energies, measured relative to the $s$ mode, of
$E_{3}=9.2k_{B}T$, $E_{barrier} =16.12k_{B}T$ and
$E_{r}=-4.6k_{B}T$. Lines corresponds to Eq. \ref{2exp} with
calculated values of $\tau_{1}$, $\tau_{2}$ and $q$. Note that here $\lambda_{u}$ is different from Fig.
\ref{fig2}.}
\label{fig3}%
\end{figure}

To gain an understanding of the difference between the two time scales
$t^{typ}_{n_{p}}$ and $t^{ave}/n_{p}$ we first consider $n_{p}=1$ in a
simplified model where $\lambda^{i}_{r}$ and $\lambda^{i}_{s}$ are independent
of $i$ except at the target site $\mathcal{T}$ where $\lambda^{\mathcal{T}%
}_{r}=\infty$ and $\lambda^{\mathcal{T}}_{s}=0$. The disorder of the DNA
sequence is neglected and the target is designed such that a reaction
takes place at the first visit of the target site. As stated above, we are
interested in the probability $\mathcal{R}(t)=\int_{0}^{t} P(t^{\prime})
dt^{\prime}$ that a reaction occurs before time $t$, where $P(t)$ is the
distribution of the first-passage time (FPT)
\cite{RednerBook,nature2007,Loverdo2008nx} to the target (we drop the
subscript when $n_{p}=1$).

The Laplace transform, ${\tilde P}(s)=\int_{0}^{\infty} e^{-st}
P(t) dt$, of $P(t)$ can be obtained exactly. To do
this we consider a DNA molecule of $N$ sites. For simplicity we take
a centered target site (labeled $0$). Consider, first, the joint
probability density for a protein to find the target at time
$t=t_{s}+t_{r}$ starting from a location $x_{0}$ at $t=0$ before
unbinding from DNA. Here $t_{s}$ is the total time spent in the $s$
state and $t_{r}$ is the total time spent in the $r$ state. If
exactly $n$ transitions occurred from the $s$-state to the $r$-state
this is given by
\begin{equation}
P_{n}(t_{s},t_{r}|x_{0}) =\lambda_{s} \mathrm{\mathcal{P}}
(n-1,\lambda_s,t_r)
\mathrm{\mathcal{P}}(n,\lambda_{r},t_{s})j(t_{s}|x_{0})e^{-\lambda_{u}t_{s}
},\label{psd}
\end{equation}
where $\mathrm{\mathcal{P}}(n,\mu,t)=(\mu t)^{n}e^{-\mu t}/n!$ is
the Poisson distribution and we use the convention
$\mathrm{\mathcal{P}}(-1,\mu,t)\equiv \delta(t)/\mu$. $j(t|x_{0})$
is the FPT density at the target $x=0$ for a usual random walk
starting from $x_{0}$ whose functional form was derived in
\cite{montroll}. The FPT density before unbinding starting from
$x_{0}$ then reads:
\begin{equation}
\displaystyle J(t|x_{0})=\sum_{n=0}^{\infty}\int_{0}^{\infty}\!\!\int
_{0}^{\infty}\!\!dt_{s}dt_{r}\delta(t_{s}+t_{r}-t)P_{n}(t_{s},t_{r}%
|x_{0}).\label{psd2}%
\end{equation}
After Laplace transform and using
${\tilde{\mathrm{\mathcal{P}}}}(n,\mu,s)=\mu ^{n}/(s+\mu)^{n+1}$, we
find ${\tilde{J}}(s|x_{0})={\tilde{j}}(u(s)|x_{0})$ with
$u(s)=\frac{s(s+\lambda_{r}+\lambda_{s}+\lambda_{u})+\lambda_{s}\lambda_{u}
}{s+\lambda_{s}}$. Averaging over $x_{0}$ and  following
\cite{nousprotein,pccp2008} we finally obtain
\[
\displaystyle\tilde{P}(s)=\tilde{j}(u(s))\left\{  1-\frac{\lambda_{b}%
\lambda_{u}}{s+\lambda_{b}}\frac{1-\tilde{j}(u(s))}{u(s)}\right\}
^{-1},
\]
where
$\tilde{j}(s)\equiv\left\langle \tilde {j}(s|x)\right\rangle
_{x}\sim\frac{1}{N}\sqrt{\frac{1+e^{-s/\lambda_{0}}}%
{1-e^{-s/\lambda_{0}}}}$ for large $N$ \cite{montroll}.

The results along with numerics, performed using a
standard continuous time Gillespie algorithm, are shown in Fig. \ref{fig2}. As is
clearly evident, for a realistic range of parameters (we take
barrier heights to be of the same order of magnitude as the
experimentally measured binding energies)
$\mathcal{R}(t)$ reaches a plateau close to one on a typical time scale
$t^{typ}$ which, for $N=10^{6}$, is much shorter than the average search time
$t^{ave}=-\frac{d{\tilde{P}}}{ds}(s=0)$. Quantitatively the typical search time
$t^{typ}$ can be defined, for example, through the median $\mathcal{R}(
t^{typ})=1/2$. For analytical purposes, we find it useful to define it
through
\begin{equation}
\int_{0}^{\infty}e^{-t/t^{typ}}P(t) dt = {\tilde
P}(1/t^{typ})=1/2\;.
\label{ttyp}%
\end{equation}
Experimentally, the relevant time, where almost all search processes
end, is $t^{typ}$ and not $t^{ave}$.

\begin{figure}[ptb]
\begin{center}
\includegraphics[width=6cm ]{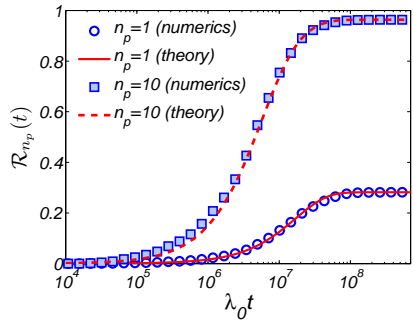}
\end{center}
\caption{Plot of $\mathcal{R}_{n_{p}}(t)$ for $n_{p}=1$ (empty
circles) and $n_{p}=10$ (filled squares) for the disordered model.
The lines were obtained by fitting the form
$1-(qe^{-t/\tau_{1}}+(1-q))^{{n}_{p}}$ to the numerical simulations with $q=0.2817$,
$\lambda_0\tau_{1}=1.7\cdot10^{7}$ and $\tau_{2}=\infty$. These are close to
the mean field prediction $q=0.2827$,
$\lambda_0\tau_{1}=1.1\cdot10^{7}$. Here $\lambda_{3}=10^{-2}\lambda_0$ ($E_{3}=4.6k_{B}%
T$), $\lambda_{b}=0.1\lambda_0$, $E_{0}=30k_{B} T$ and
$\sigma=5.3k_{B}T$. Note that here the average height of
the barrier at the target site is $6.25k_{B}T$.}
\label{fig4}%
\end{figure}

Importantly, the distribution $\mathcal{R}(t)$ has two intrinsic time scales,
one short and one long, and can in practice be well approximated by
\begin{equation}
\mathcal{R}(t) \simeq1-qe^{-t/\tau_{1}}-(1-q)e^{-t/\tau_{2}} \label{2exp}%
\end{equation}
where $q$, $\tau_{1}$ and $\tau_{2}$ can be calculated analytically. This form allows an explicit determination of $t^{typ}$ (through Eq. (\ref{ttyp})) and enables
the following interpretation. The short time scale $\tau_{1}=-\frac{1}{q}\frac{d{\tilde{P}}}{ds}(\lambda_{s}=0,s=0)$ characterizes
events where the protein never enters the $r$ state and is therefore
independent of the binding energy $E_{r}$ (and hence of $\lambda_{s}$) ; $q=\tilde{P}(\lambda_{s}=0,s=0)$
is the probability of such an event. The time scale $\tau_{2}=(t^{ave}-q\tau_{1})/(1-q)$ characterizes
events where the protein enters the $r$ state, and is therefore much larger
than $\tau_{1}$ in the case of strong binding ($\lambda_{s}$ small). As
illustrated in Fig. \ref{fig2} the competition between the two time scales can
lead, for DNA lengths which are experimentally relevant, to a significant
difference between the typical
and average times. More precisely, we find that for DNA lengths $N< \sqrt{2\lambda_{0}\lambda_{u}}/\lambda
_{r}$, $q$ is of the order of one and $t^{typ}\simeq\tau_{1}\simeq
N\sqrt{\frac{\lambda_{u}}{2\lambda_{0}}}(\lambda_{u}^{-1}+\lambda_{b}^{-1})$
is independent of $\lambda_{s}$ - the only rate which depends on the binding
energy in $r$ mode. The relevant time scale of the search process
$t^{typ}$ can therefore be much shorter than $t^{ave}\simeq N\lambda_{r}
/\lambda_{s}\sqrt{2\lambda_{0}\lambda_{u}}$ even in the presence of deep traps
($\lambda_{s}$ small).

This interesting regime where $t^{typ}\ll t^{ave}$ requires a rather large
barrier between the $s$ and $r$ state in the case of long DNA molecules
(namely, $\lambda_{r}< \sqrt{2\lambda_{0}\lambda_{u}}/ N$). We now argue that
this constraint can be, to a large extent, relaxed when $n_{p}$ proteins are
searching for the target simultaneously. In this case even when for a single
protein $t^{ave}\cong t^{typ}$ the typical search time $t^{typ}_{n_{p}}$ of
$n_{p}$ proteins can be significantly shorter than $t^{ave}/n_{p}$ even for
relatively small values $n_{p} \approx10 - 15$. Here, again, $t^{ave}$ is the
average search time of a single protein and $t^{typ}_{n_{p}}$ is defined as in
Eq. \ref{ttyp} where for $n_{p}$ proteins the first-passage distribution
$P_{n_{p}}(t)$ is deduced from the cumulative distribution
\begin{equation}
\mathcal{R}_{n_{p}}(t)=1-\left(  1- \mathcal{R}(t) \right)  ^{n_{p}} \;.
\label{Pnp}%
\end{equation}
In Fig. \ref{fig3} we show the results of $\mathcal{R}_{n_{p}}(t)$
for $n_{p}=10$. Note that as claimed above $t^{typ}_{n_{p}} \ll
t^{ave}/n_{p}$, whereas $t^{typ}$ is close to $t^{ave}$ for one
protein. This can be understood in the following manner. Using the
approximate form, Eq. \ref{2exp}, in Eq. \ref{Pnp}, it is obvious
that when $\tau_{2}\gg\tau_{1}$, the decay of
$\mathcal{R}_{n_{p}}(t)$ is dominated by $\tau_{1}$ as long as
$(1-q)^{n_{p}} \ll1$. In essence since only one protein needs to
find the target, the probability of a catastrophic event where the
search time is of the order of $\tau_{2}$ is $p_{cat}=(1-q)^{n_{p}}$
which decays exponentially fast with $n_{p}$. For large enough
values of $n_{p}$ the short time scale $\tau_{1}$ controls the
behavior of $\mathcal{R}_{n_{p}}(t)$, even if it is insignificant
for the one protein search time.
This implies that searches involving several proteins strongly suppress the
long time-scales induced by the traps which control $t^{ave}$.
The typical search time is then given by $t^{typ}_{n_{p}} = \tau_{1}/m$, where
$m$ is of the order of $n_{p}$, and is therefore again widely independent of
the binding energy of the $r$ mode. This makes fast searches possible even in
the presence of deep traps - enabling both speed and stability.

We now argue that this mechanism of fast search can still be at play when the
binding energy of the protein to the DNA is strongly disordered, as observed
in experiments. To account for this we consider the case where
the barrier height is drawn from a Gaussian distribution: $p(E^{i}%
_{b})=e^{-(E^{i}_{b}-E_{0})^{2}/2
\sigma^{2}}/\sqrt{2\pi\sigma^{2}}$. Importantly, in the presence of
disorder we can propose an intrinsic definition of the target as the
site with the \textit{lowest barrier} with no specifically designed
properties. Indeed, our previous assumption $\lambda^{\mathcal{T} }_{r}=\infty$ at
the target site and $\lambda_{r}^{i}$ small everywhere else is a
rather strong demand. Since the target sequence is of the order of
10 base-pairs, many sequences with similar properties are very
likely to exist, unless the DNA sequence is carefully tailored.
\begin{figure}[ptb]
\begin{center}
\includegraphics[width=8cm]{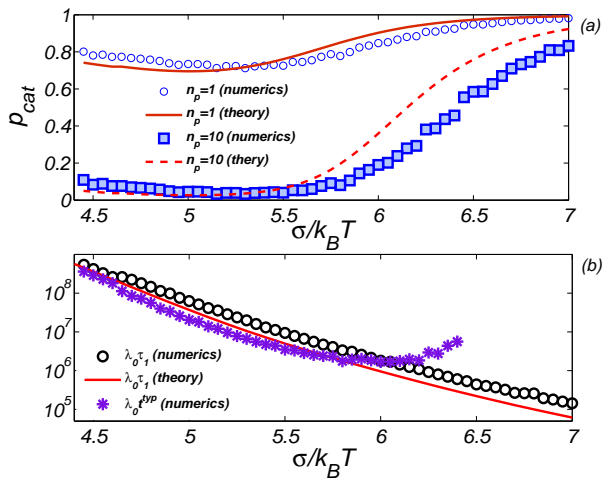}
\end{center}
\caption{Results for the disordered model. Here $N=10^{6}$,
$\lambda_{3}=10^{-2}\lambda_0$ ($E_{3}=4.6k_{B}T$),
$\lambda_{b}=0.1\lambda_0$ and $E_{0}=30k_{B}T$. (a) $p_{cat}$ as a
function of $\sigma$ for $n_{p}=1$ and $n_{p}=10$. (b) $t^{typ}$ for
$n_{p}=10$ and $\tau_{1}$ are plotted as a function of $\sigma$.
Using $\lambda_{0}=10^{6} \sec^{-1}$ \cite{WAC2006} for $n_{p}=10$
at the minimal $p_{cat}$ we find $t^{typ}\simeq10sec$.}
\label{fig5}%
\end{figure}
To analyze this model we combine numerics with a mean-field analysis.
For simplicity, we consider the extreme case where all
recognition sites are infinitely long lived $\lambda_{s}=0$ (or equivalently
$\tau_{2}=\infty$), which obviously fulfills the stability requirement. Note
that the average search time is then infinite.

Within the mean field approach we replace the different quantities
by their disorder average and account for the barrier at the target
site. We first compute the disorder averaged probability of crossing
the barrier at the target at each visit. Knowing the distribution of
the minimum of the barrier \cite{HF2006}, this is given by
$p_{1}=\int_{-\infty}^{\infty}dE \frac{e^{-E/k_B T}
}{1+\lambda_{u}/\lambda_0+e^{-E/k_B T}}\frac{d}{dE}\left[
\frac{1}{2}\operatorname{erfc} \left(
\frac{E-E_{0}}{\sqrt{2}\sigma}\right)  \right]  ^{N}$. Here we set
the time scale of the activation process across the barrier to be
$\lambda_0$. We finally assume that the expression for $u(s)$ of the
non-disordered model holds with $\lambda_{r}$ replaced by
${\bar{\lambda}}_{r}=\lambda_{0}\int_{-\infty}^{\infty} e^{-E/k_B T}
\frac{e^ {-\frac{\left(E- E_{0}\right) ^{2}}{2\sigma^{2}}}}
{\sqrt{2\pi}\sigma}dE$ and $\tilde{j}$ replaced by
\begin{equation}
\tilde{j}_{p_{1}}=\frac{p_{1}{\tilde{j}}(z)}{1-(1-p_{1}){\tilde{j}}_{0}(z)}
\label{psd3}
\end{equation}
where ${\tilde{j}}_{0}(s)$ is the
generating function of the first return time to site 0
\cite{montroll}.

First, we show that the two scales scenario described above still holds.
Indeed, Fig. \ref{fig4} shows that $\mathcal{R}(t)$ is well fitted by Eq.
\ref{2exp} for realistic values of parameters. This implies that for $n_{p}$
large enough the only relevant time scale is $\tau_{1}$ and the typical search
time again takes the form $t^{typ}_{n_{p}} \simeq\tau_{1}/m$ with $m$ of the
order of $n_{p}$. This enables a fast search even in the presence of
infinitely deep traps.

The regime of a fast search with $t^{typ}_{n_{p}}$ independent of the trap
depth $E_{r}$ also requires, as above, a small $p_{cat}$. We now show that
this condition holds in a wide range of disorder parameters. To illustrate
this, the dependence (holding all other variables constant) of $p_{cat}$ and
$t^{typ}_{n_{p}}$ on $\sigma$, obtain from numerics and the mean-field treatment, is shown in Fig. \ref{fig5} for realistic values
of parameters. Notably, the value of $p_{cat}$ can be minimized as a function
of $\sigma$. This reflects the fact that for small values of $\sigma$ the DNA
sequence has to be scanned many times before the target is entered in the
$r$-mode. Increasing $\sigma$ lowers the barrier at the target and therefore
reduces the number of scans needed, which diminishes $p_{cat}$. For larger
$\sigma$ the chance of falling into a trap increases due to lower secondary
minima of the barrier, which leads to an increase of $p_{cat}$.
As expected, $p_{cat}$ is dramatically decreased when $n_{p}$ is increased,
even by a few units, and can remain small for a wide range of values of
$\sigma$. For larger $\sigma$, $p_{cat}$ increases and $t^{typ}_{n_{p}}$ rises quickly
 as it starts to depend on $\tau_{2}$.

Most important, as advertised above, these results show that it is possible to
obtain relatively small values of $t^{typ}_{n_{p}}$ and $p_{cat}$ with
realistic values of the parameters (see Fig. \ref{fig5}). Reasonable search
times (in the range of seconds) are obtained for a rather large range of
$\sigma$ as long as $n_{p}$ is of the order of ten or more proteins, even in
the extreme case of infinitely deep traps suggesting a possible resolution of the speed and stability requirements.
We note that by moderate changes in $E_{0}$ similar
results can be obtained for much longer DNA sequences.

We thank E. Braun and D. Levine for comments and the support of the High Council for Scientific and Technological
Cooperation between France-Israel. Y. K. and M. S.
were also supported by the Israeli Science Foundation, and O.B. and R.V. by ANR grant "Dyoptri".

\end{document}